\documentclass[preprint]{aastex}

\begin{document}

\title{Landau Damping and Alfven Eigenmodes of Neutron Star Torsion Oscillations} 

\author{Andrei Gruzinov}

\affil{CCPP, Physics Department, New York University, 4 Washington Place, New York, NY 10003}

\begin{abstract}

Torsion oscillations of the neutron star crust are Landau damped by the Alfven continuum in the bulk. For strong magnetic fields (in magnetars), undamped Alfven eigenmodes appear.

\end{abstract}

\section{Introduction and Conclusion}

It is thought that Israel et al (2005) have detected torsion oscillations of the neutron star crust. Here we show that crustal modes are strongly affected by the Alfven continuum in the bulk. For week magnetic fields, there is Landau damping of crustal eigenmodes. At stronger magnetar-like magnetic fields undamped Alfven eigenmodes appear. In this regime, the Alfven waves in the bulk control the torsion oscillations of the crust.

It might be that Israel et al (2005) have actually measured the Alfven speed in the bulk $c_A$, rather than the torsion speed in the crust $c_t$. If so, the measured $\nu \approx 20$Hz frequency gives $c_A \approx 2\nu R \approx 4\times 10^7$cm/s (assuming that $\nu$ is the fundamental Alfven eigenmode).

We  calculate the Landau damping of the crustal torsion modes and Alfven eigenmode (undamped) frequencies. The thin-crust uniform-field constant-density model is used, and we only consider the antisymmetric axisymmetric modes. This is obviously not good enough for serious neutron star seismology. One must include: (i) the non-axisymmetric modes, (ii) the toroidal magnetic field, (iii) the realistic density profile. But the qualitative results -- Landau damping and emergence of undamped Alfven eigenmodes -- should be robust.

This paper is a formal mathematical development of the ideas of Levin (2007).

\section{Results}

We use the following units and notations: 

\begin{itemize}

\item $R=1$ is the stellar radius

\item $c_A=1$ is the Alfven speed in the bulk

\item $3M$ is the crust to bulk mass ratio. 

\item $c_t$ is the torsion sound speed in the crust (in units of $c_A$).

\end{itemize}

\begin{table}
\begin{center}
\caption{Eigenfrequencies in units of $c_A/R$.}

\begin{tabular}{cr}

\tableline
\tableline

$M=0.1$, $c_t=1$ & 2.10, 2.74, 3.03, 3.12, 5.05-0.001i, 5.75-0.002i, ...  \\

\tableline

$M=0.1$, $c_t=3$ &  3.06, 6.18-0.11i, 9.35-0.04i, ...\\

\tableline

$M=0.1$, $c_t=10$ &  21.7-4.01i, 46.3-6.19i, ...\\

\tableline

$M=0.1$, $c_t=100$ &  200-3.14i, 424-3.19i, ...  \\

\tableline

\tableline

\tableline

$M=0.01$, $c_t=10$ & 3.11, 6.26-0.005i, 9.41-0.006i,  ...   \\

\tableline

$M=0.01$, $c_t=30$ & 103-76i, 198-144i,  ...   \\

\tableline

$M=0.01$, $c_t=100$ & 219-55i, 490-102i,  ...   \\

\tableline
\tableline

\end{tabular}

\end{center}
\end{table}

The calculated frequencies $\omega$ (in units of $c_A/R$) of the antisymmetric axisymmetric torsion modes ($l=2,4,6,...$, $m=0$) are given in the table for various mass ratios $M$ and speed ratios $c_t$. The eigenmodes (and cuts, see \S3 ) are also shown in the figure for $M=0.1$, $c_t=2$.

\section {The eigenmode equation and the dispersion relation}

Toroidal displacements of the crust launch Alfven waves into the bulk. For an axisymmetric antisymmetric torsion mode, the toroidal displacement in the bulk is $\propto e^{-i\omega t}\sin (\omega z)$, where $z$ is the Cartesian coordinate along the magnetic field, with the equatorial plane $z=0$. Including the corresponding force into the thin-crust eigenmode equation, we get
\begin{equation}
c_t^2\left( (1-x^2)f''-4xf'\right) +\omega ^2f  ={\omega x \cot (\omega x)\over M}f.
\end{equation}
Here $x\equiv \cos \theta$, $\theta$ is the spherical angle with $\theta =0$ along $z$, $f\sin \theta$ is the toroidal displacement of the crust, prime is the x-differentiation. The detailed derivation is given in the Appendix. 

For the antisymmetric mode, the boundary condition at $x=0$ is $f=0$. The boundary condition at $x=1$ is given by the equation itself: $-4c_t^2f'+\omega ^2f  ={\omega \cot (\omega )\over M}f$. 

To obtain the dispersion relation in the form $F(\omega )=0$, choose some $f(1)$ and $f'(1)$ satisfying the $x=1$ boundary condition. Then integrate (1) form $x=1$ to $x=0$ and put $F(\omega )=f(0)$. 

According to the Landau rule, one calculates $F(\omega )$ in the upper half-plane of complex $\omega$. The eigenmodes are given by the zeros of  $F(\omega )$ on the real axis (undamped modes) and in the lower half plane (Landau damped modes). The value of $F$ in the lower half-plane should be obtained by the analytic continuation from the upper half-plane. 

The analytic continuation of $F$ to the lower half-plane can be accomplished by complexifying $x$ and integrating (1) from $x=1$ to $x=0$ along an arbitrary contour in the upper half-plane of $x$. The shape of the x-contour determines the shape of the cuts in the lower half-plane of $\omega$. We chose the semicircle $|x-0.5|=0.5$. This gives the cuts in the lower half-plane of $\omega$ running straight down from $\omega = \pi k$. 

The dispersion function $F$ was calculated numerically. The eigenmodes, both the Landau-damped crustal modes and the undamped Alfven eigenmodes are given in the table. The real entries of the table (what we call Alfven eigenmodes), were confirmed by a straightforward real-numbers integration of (1). 

Due to the cuts, the late-time asymptotic of the crustal motion will also have algebraically damped modes with frequencies $\omega = \pi k$. The late time asymptotic is determined solely by the location of the tips of the cuts and is not affected by the choice of the x-contour. The zeros of $F$ are also independent of the x-contour.

For generic parameters $M$ and $c_t$, numerical integration seems to be the only way. But there are limiting cases which can be treated analytically. These may serve to confirm that equation (1) actually makes sensible predictions and also to check the numerical results:

\begin{itemize}

\item $c_t\ll 1$, $M\gg 1$ gives undamped crustal modes $\omega ^2=(l^2+l-2)c_t^2+{1\over M}$, $l=2,4,6,...$. This case is obvious -- magnetic field and the fluid bulk just follow the slow and heavy crust, providing an additional elasticity and therefore increasing the frequency. This case is unphysical, because the crust is actually lighter than the bulk.

\item $c_t\gg 1$, $Mc_t\gg 1$ gives Landau damped crustal modes $\omega =\sqrt{l^2+l-2}~c_t-i{a_l\over M}$, $l=2,4,6,...$, $a_l$ are calculable dimensionless numbers (of order unity for low $l$). This case is physical, it occurs for small enough magnetic fields.

The calculation is as follows. By the Euler's formula, 
\begin{equation}
\cot (\omega x)=\sum _k {1\over \omega x -\pi k}\rightarrow \int dk {1\over \omega x -\pi k}=-i.
\end{equation}
Now (1) becomes
\begin{equation}
c_t^2\left( (1-x^2)f''-4xf'\right) +\omega ^2f  =-i{\omega x \over M}f.
\end{equation}
We solved it using the first order perturbation theory in $1/(Mc_t)$. For $l=2$, one calculates $a_2=5/16=0.313$ -- in agreement with the $M=0.1$, $c_t=100$ entry of the table.

\end{itemize}

\acknowledgements

I thank Yuri Levin for showing me the problem and for useful discussions.  

This work was supported by the David and Lucile Packard foundation. 

\appendix

\section{Torsion oscillations of the elastic crust with the magnetized bulk}

Consider a neutron star with a thin crust. Choose the $z$ axis along the uniform magnetic field $B$. The toroidal displacement of the fluid bulk $\xi$ is described by the Alfven wave equation 
\begin{equation}
\partial _t^2\xi=c_A^2\partial _z^2\xi,
\end{equation}
where $c_A^2={B^2\over 4\pi \rho }$ is the Alfven speed in the bulk of density $\rho$. 

The toroidal displacement of the crust $\psi$ is described by the torsion wave equation with the magnetic force from the bulk
\begin{equation}
\partial _t^2\psi ={c_t^2\over R^2} \left( \partial _\theta ^2\psi +{\cos \theta \over \sin \theta }\partial _\theta \psi -{1\over \sin ^2\theta }\psi +2\psi \right) + {\cos \theta T_{\phi z}\over \sigma }.
\end{equation}
Here $c_t$ is the torsion sound speed in the crust, $R$ is the radius of the star, $\theta$ is the polar angle with respect to $z$,  $T_{\phi z}$ is the toroidal-vertical component of the Maxwell stress tensor, $\sigma$ is the surface density of the crust. 

The Maxwell stress is given by $T_{\phi z}={-BB_\phi \over 4\pi }$, $B_\phi =B\partial _z\xi$, which should be calculated at the boundary of the bulk, where $\xi = \psi$. We get:

\begin{equation}
\partial _t^2\psi ={c_t^2\over R^2} \left( \partial _\theta ^2\psi +{\cos \theta \over \sin \theta }\partial _\theta \psi -{1\over \sin ^2\theta }\psi +2\psi \right) - {\rho c_A^2\over \sigma}\cos \theta \partial _z\xi |_{z=cos \theta}.
\end{equation}

As a check, one confirms that this system conserves energy and angular momentum:
\begin{equation}
E={1\over 2}\int \rho dV~\left( (\partial _t\xi )^2+c_A^2(\partial _z\xi )^2\right) + {1\over 2}\int \sigma dA~\left( (\partial _t\psi )^2+{c_t^2\over R^2}(\sin \theta \partial _\theta ({\psi \over \sin \theta }))^2\right)
\end{equation}
\begin{equation}
L=\int \rho dV ~r\sin \theta ~\partial _t \xi + \int \sigma dA ~R\sin \theta ~\partial _t\psi ,
\end{equation}
where $\int dV$ and $\int dA$ are volume and surface integrals and $r$ is the radius.

\begin{figure}
\includegraphics[angle=0,scale=0.8]{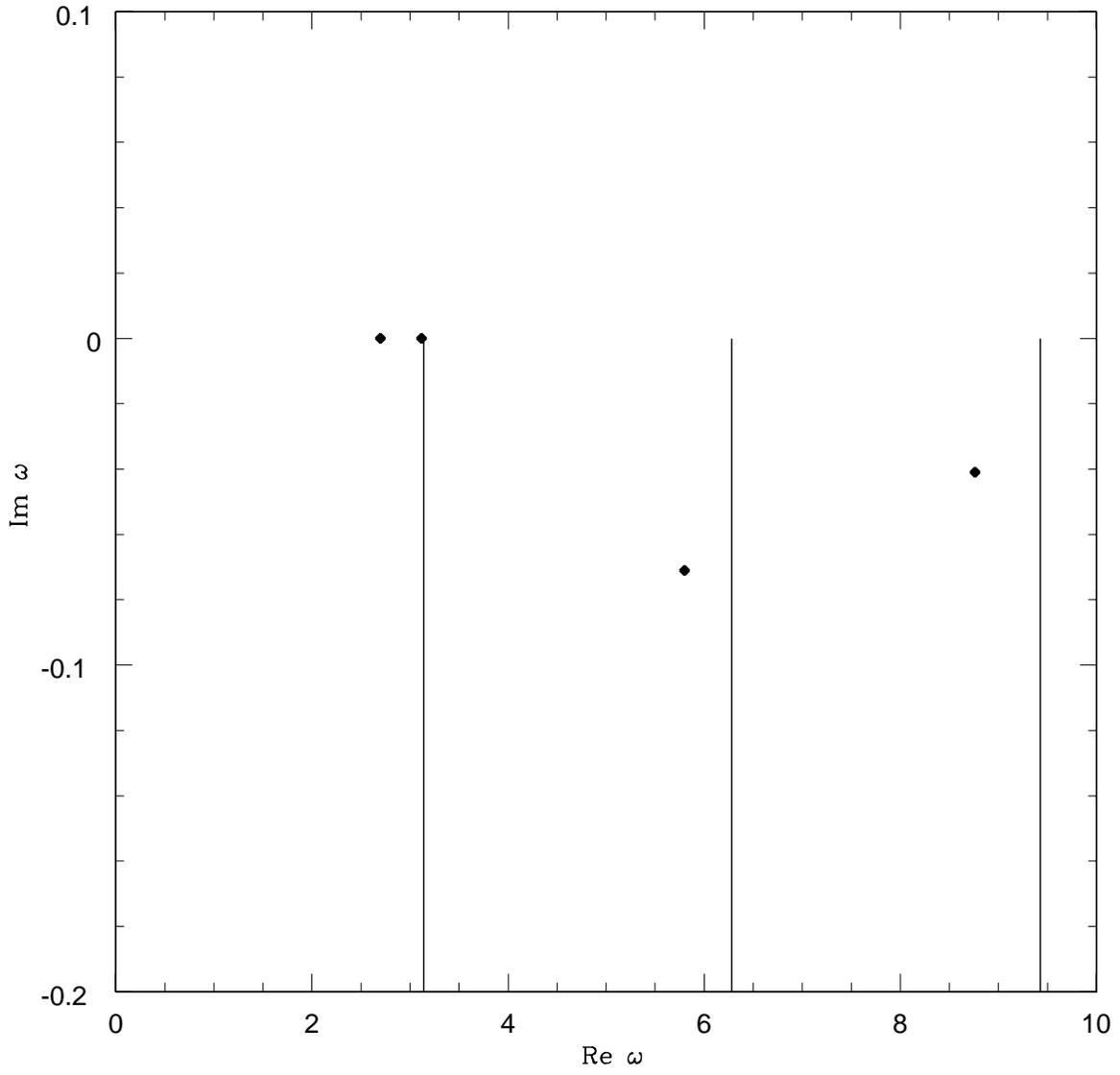}
\caption{ $M=0.1$, $c_t=2$. Zeros and cuts of the dispersion function $F(\omega )$ in the complex $\omega$ plane. }
\end{figure}

\end{document}